\documentclass[preprint,aps,nofootinbib]{revtex4}
\usepackage{graphicx}
\usepackage{color}

\def\bold#1{\setbox0=\hbox{$#1$}%
     \kern-.025em\copy0\kern- -\wd0
     \kern.05em\copy0\kern-\wd0
     \kern-.025em\raise.0433em\box0 }

\newcommand{\gev}{\,\mbox{GeV}}

\def\ga{\mathrel{\raise.3ex\hbox{$>$\kern-.75em\lower1ex\hbox{$\sim$}}}}
\def\la{\mathrel{\raise.3ex\hbox{$<$\kern-.75em\lower1ex\hbox{$\sim$}}}}

\def\m12{m_{1\!/2}}

\def\lsim{\mathrel{\raise.3ex\hbox{$<$\kern-.75em\lower1ex\hbox{$\sim$}}}}
\def\gsim{\mathrel{\raise.3ex\hbox{$>$\kern-.75em\lower1ex\hbox{$\sim$}}}}

\begin{document}
\begin{titlepage}
\pagestyle{empty}
\baselineskip=21pt
\rightline{UCD--2005--07}
\vskip 1in
\begin{center}
{\large{\bf  \boldmath
Invisible Quarkonium Decays as a Sensitive Probe of Dark Matter
}}
\end{center}
\begin{center}
\vskip 0.2in 
{\bf  Bob McElrath}\footnote{mcelrath@physics.ucdavis.edu}
\vskip 0.1in 
{\it Department of Physics, University of California,
Davis, CA 95616}

\vskip 0.2in {\bf Abstract}
\end{center}
\baselineskip=18pt
\noindent 
We examine in a model-independent manner the measurements that can be
performed at B-factories with sensitivity to dark matter.  If a singlet
scalar, pseudo-scalar, or vector is present and mediates the Standard
Model - dark matter interaction, it can mediate invisible decays of
quarkonium states such as the $\Upsilon$, $J/\Psi$, and $\eta$.
Such scenarios have arisen in the context of supersymmetry, extended
Higgs sectors, solutions the supersymmetric $\mu$ problem, and extra
$U(1)$ gauge groups from grand unified theories and string theory.
Existing B-factories running at the $\Upsilon(4S)$ can produce lower
$\Upsilon$ resonances by emitting an Initial State Radiation (ISR)
photon.  Using a combination of ISR and radiative decays, the initial
state of an invisibly decaying quarkonium resonance can be tagged,
giving sensitivity to the spin and CP-nature of the particle that
mediates standard model-dark matter interactions.  These measurements
can discover or place strong constraints on dark matter scenarios where
the dark matter is approximately lighter than the $b$-quark.  For the
decay chains $\Upsilon(nS) \rightarrow \pi^+ \pi^- \Upsilon(1S)$ (n=2,3)
we analyze the dominant backgrounds and determine that with $400
fb^{-1}$ collected at the $\Upsilon(4S)$, the B-factories can limit
$BR(\Upsilon(1S) \rightarrow invisible) \lsim 0.1\%$.

\end{titlepage}
\baselineskip=18pt
%%%%%%%%%%%%%%%%%%%%%%%%%%%%%%%%%%%%%%%%%%%%%%%%%%%%%%%%%%%%%%%%%%%%%
\setcounter{footnote}{0}
\section{Introduction}

Multiple astrophysical experiments have measured the presence of Dark
Matter (DM).  The leading candidate for this DM is a
particle.~\cite{dmreview}  Existing studies have concentrated on a single
particle providing the DM density of the universe but
multiple particles are allowed.  This knowledge should lead to a
systematic search for invisible decays of particles and mesons known to
exist.  However, the only particles with reported invisible branching
ratios or limits are the $\pi^0$ and $Z$.~\cite{pdg}

There are at least two pieces of evidence that the DM component
of the universe may be lighter than the Minimal Supersymmetric Standard
Model (MSSM) or minimal supergravity mediated supersymmetry breaking
(mSUGRA) lightest
neutralino that dominates DM studies.  First, recent
measurements of 511 keV gamma rays from the galactic
center indicate a Gaussian profile of low-velocity
positrons.~\cite{Jean:2003ci}  Traditional MSSM or mSUGRA neutralinos are heavy, and their
annihilation would produce too many high-energy $\gamma$-rays from 
neutral pions which decay to photons, as well
as significant bremsstrahlung as their decay products slow until they are
nearly at rest as is required explain the 511 keV line.  This places such a scenario
in strong conflict with EGRET upper limits on the higher-energy gamma
flux.~\cite{Strong:2004de,Beacom:2004pe}  Thus, if a DM particle is
responsible for the 511 keV line, it must be lighter than approximately
100 MeV.  Second, recent analysis of DM flows and caustics indicate that
the CDMS limit and DAMA evidence for DM can be compatible due to the
lower detection threshold of DAMA.~\cite{dmflows}  This effect is also
enhanced if there is a flow of DM through our solar system.

Neutralinos
in the general MSSM must have 
$M_{\chi^0} > 6$ GeV to obtain an appropriate relic
density.~\cite{Bottino:2002ry}  Constraints on even lighter DM comes 
from CUSB, which measured $\Upsilon \rightarrow \gamma + {\rm
invisible}$ signals, giving the best sensitivity to DM lighter than approximately
$1.5 \gev$, and losing sensitivity as the DM mass increases due to the
soft spectrum of Initial State Radiation (ISR)
photons.~\cite{ups1stogammainvis}  Modern b-factories can improve in
this measurement by at least an order of magnitude.\cite{mySinglinoDM}
Searches for invisibly-decaying Higgs
Bosons are sensitive if the Higgs is heavy, has significant coupling to
the $Z$, and the Higgs decays dominantly to dark
matter.~\cite{invisiblehiggs} Finally, the LEP single-photon counting
measurements limit arbitrary Standard Model (SM)-DM interactions; however, the $Z$
invisible width dominates these measurements at LEP energies, and these
experiments have no sensitivity if the SM-DM mediator does not couple to
the electron.~\cite{lepsinglephoton}

These constraints can all be avoided in many models that are not the
MSSM.  Two attractive possibilities that can explain the above data are
(1) a light neutralino with couplings to the SM mediated by
a light scalar singlet.  This may occur in the Next-to-Minimal
Supersymmetric Model (NMSSM) and
related models which solve the $\mu$ problem with a
singlet;~\cite{nmssm,mySinglinoDM}
(2) light scalar DM coupled to the Standard Model through a
new gauge boson $U$.~\cite{Boehm:2003hm,morelightdm,Fayet:1991ux}
In both cases there must be some small coupling to the Standard Model in
order to avoid having the DM density over-close the
universe.~\cite{dmreview}
It is in general possible 
to couple the DM preferentially to some quarks and/or leptons but not
others.
This can result in invisible decays of some hadrons of a given spin $J$ and
Charge$\times$Parity (CP) eigenvalue
but not others, and little or no signal at
direct detection experiments.  Building such models is
straightforward, and several already exist in the
literature.~\cite{nmssm,Boehm:2003hm}  Our
purpose in this letter is not to build such models, but point out
several measurements that can be performed at colliders that
are sensitive to light DM.

At $e^+ e^-$ detectors like BaBar, Belle, and CLEO, one can use ISR
to explore energy regions below the nominal collider
energy.~\cite{isrtheory,isrexpt}  With one ISR photon, these
experiments
deliver the same luminosity between 9.85 GeV and 10.58 GeV as they do
from interactions without an ISR photon at their nominal center of mass
energy, 10.58 GeV.  In order to identify the presence of a bottomonium
state without observing its decay, one must require an ISR photon
of a specific energy and/or a radiative decay.  A radiative decay is any
transition from one quarkonium state to another.
We present
several techniques which can be used to suppress backgrounds when the
ISR photon is lost because it is outside the detector acceptance.
Radiative decays from the $\Upsilon(4S)$ will have similar statistics to
ISR production of lower $\Upsilon$ resonances, but no radiative decays
of the $\Upsilon(4S)$ have yet been discovered.

Another measurement that can be performed at B-factories is in $b
\rightarrow s$ transitions such as $B^+ \rightarrow K^+ + invisible$ and
is sensitive to dark matter with masses up to 2.4 GeV, but is not
sensitive to the $J^{CP}$ of the mediator.\cite{bsme}  This branching
ratio may be 50 times larger than is expected from the Standard Model
process with neutrinos.

The Standard Model expectation for $\Upsilon$ to invisible is
$\Gamma(\Upsilon \rightarrow \nu \bar{\nu}) = 4.14\times 10^{-4}
\Gamma(\Upsilon \rightarrow e^+ e^-) \simeq 1\times 10^{-5}$ with a
theoretical uncertainty of only $2-3\%$, and is sensitive to the bottom
squark mass and R-parity violation in SUSY theories.~\cite{sminvisible}  

Expectations for branching ratios of hadrons into DM may be as large as
a few percent.

\section{Sources of Dark Matter}
We assume that DM couples to the Standard Model through some
mediating boson.  On general model-independent grounds we expect this
particle to be either a vector, scalar, or pseudo-scalar.  If the
mediator is a scalar or pseudo-scalar, only a 
$SU(2)$ doublet can couple to $b \bar{b}$ by gauge invariance at
dimension 4.  This scalar doublet will
generally mix with the Standard Model Higgs, or the CP-odd $A$ of a Two
Higgs Doublet Model.  Therefore, we expect scalar and pseudo-scalar
mediated DM to show up dominantly in interactions with heavy
fermions such as $b$ quarks.

If the mediator is a vector gauge boson, giving $b$ and $\bar{b}$ equal
and opposite charge under the new gauge group (which we assume to
be a $U(1)$) is sufficient to introduce the proper
couplings.~\cite{Fayet:1991ux,Boehm:2003hm}  One need not expect
this gauge boson to couple to all fermions in the Standard Model equally.  
One might expect that the first two generations are not charged
under this new group, in order to be consistent with precise
measurements of the muon and electron anomalous magnetic moment, as well
as the lack of unexplained vector resonances in hadronic data.  This
situation would result in extremely small DM-nucleon cross sections
for direct detection experiments.

In order to get small DM masses in the MSSM and other models, one
generally has to also bring down another particle mass for the purpose
of getting a large enough annihilation or coannihilation cross section.
However, in the case of DM masses less than $M_\Upsilon/2=4.73$ GeV that we consider, one
generally cannot bring down a particle charged under the SM gauge groups
without violating existing experimental constraints.\footnote{The usual
particles that are made light in supersymmetric models are the stau or a
Higgs.  The stau is often the next-to-lightest supersymmetric particle
and undergoes t-channel coannihilation with the LSP.  The Higgs mediates
s-channel annihilation when there is significant higgsino or wino
fraction in the LSP.}  Having a light bottom squark may be one exception
to this assumption;~\cite{Berger:2000mp} however, a recent re-analysis of
available data indicates that this solution is now
disfavored.~\cite{Janot:2004cy}

The direct constraint on the annihilation
mediator is the reason why existing constraints~\cite{Bottino:2002ry}
require $M_{\chi^0} > 6$ GeV, despite the fact that the theory is
consistent with a massless neutralino.~\cite{Gogoladze:2002xp}  If the
mediator has some mixing with a pure singlet state, these constraints
can be largely avoided.

A general argument predicts that DM should be heavier than 2
GeV.~\cite{Lee:1977ua} This is based on DM annihilation
couplings that are proportional to $G_F$.  However, there is nothing that
requires DM to have something to do with weak bosons or
electroweak symmetry breaking.  The argument that DM couplings must be
proportional to $G_F$ is based solely on the coincidence that the DM
annihilation cross section (c.f. Eq.~\ref{annxsect}) is similar in size
to weak cross sections.  This may simply be a numeric coincidence and 
annihilation cross sections need not
be proportional to $G_F$.  If we simply assume that DM exists and
$m_\chi < 6$ GeV is an allowed region, we are forced to recognize that
it must have picobarn cross sections with some Standard Model particle.
These expected cross sections are explored in the following section.

\section{Dark Matter Coupling Expectations}
The required rate of DM annihilation can be naively estimated.
We will not accurately compute the relic density since we are not
proposing a specific DM model, but one can get an order of
magnitude estimate for s-wave annihilation using~\cite{dmreview,pdg}
\begin{equation}
\Omega_X h^2 \simeq \frac{0.1 {\rm pb} \cdot c}
{\langle \sigma v \rangle } .
\end{equation}
Where $\Omega_X=\rho_X/\rho_c$ is the relic density for species $X$ relative to the
critical density $\rho_c$, $h$ is the Hubble constant, and
$\langle \sigma v \rangle$ is the thermally averaged annihilation cross
section of the DM into Standard Model particles.  
Using the central value of the
WMAP~\cite{wmap} result for $\Omega_X h^2 = 0.113$, we can invert this
equation and solve for the required annihilation cross section for light
relics
\begin{equation}
\langle \sigma v \rangle = 0.88 {\rm pb}.
\end{equation}
The velocity $v$ appearing here is the M\o ller velocity, the relative
velocity of annihilating particles at the temperature they froze-out.
The approximate temperature at freeze-out is $T=m_\chi/x_{FO}$ where
$m_\chi$ is
the mass of the DM and $x_{FO}$ is an expansion parameter
evaluated at the freeze-out temperature that is 
$x_{FO} \sim 20-25$ depending on the model.  Thus the average momentum for a
fermion is $k_B T$ and therefore the average relative velocity is
roughly $1/x_{FO}$.  For $x=20$ at freeze-out we have:
\begin{equation}
    \label{annxsect}
    \sigma(\chi \chi \rightarrow SM) \simeq 18 {\rm pb}.
\end{equation}

The invisible branching ratio of a hadron can then be estimated by
assuming that the time-reversed reaction is the same, $\sigma(f \bar{f}
\rightarrow \chi \chi) \simeq \sigma(\chi \chi \rightarrow f \bar{f})$.
This assumption holds if $m_\chi \simeq m_f$ and $M_\Upsilon \simeq 4
m_\chi^2 + 6 m_\chi T_{FO}$.
We assume that the DM mediator is not flavor changing and that annihilation occurs in the $s$
channel.~\footnote{A $t$-channel mediator is possible, but this requires
that the mediator carry color and electromagnetic charge, and therefore
is unlikely if we consider $m_\chi < 5$ GeV.}  Therefore, the
best-motivated hadrons to have an invisible width are same-flavor
quark-antiquark bound states (quarkonia).  The CERN Yellow Report
provides a thorough review of quarkonium physics.~\cite{Brambilla:2004wf}

The invisible width of a hadron composed dominantly of $q
\bar{q}$ is given approximately by:
\begin{equation}
    \Gamma(H \rightarrow \chi \chi) = f_H^2 M_H \sigma(q \bar{q}
    \rightarrow \chi \chi)
\end{equation}
where $f_H$ is the hadronic form factor for the state $H$, and $M_H$ is
the hadron's mass.

We can predict an approximate expectation for the branching ratios for
narrow states.  Some of the most promising are:
\begin{equation}
    BR(\Upsilon(1S) \rightarrow \chi \chi) \simeq 0.41\% \qquad
    BR(J/\Psi \rightarrow \chi \chi) \simeq 0.023\% \qquad
    BR(\eta \rightarrow \chi\chi) \simeq 0.033\%
\end{equation}
Branching ratios for scalars and pseudo-scalars tend to be smaller since
those states are wider.  This estimate does not take into account
kinematic factors arising from the mediator mass and DM mass.
These factors can both enhance or suppress these branching ratios.

% Referee general issue #1
If a particular hadron $H$ decays invisibly, then that hadron must mix
into the mediator $M$ before decaying into Dark Matter.  If $M$ does not
violate the discrete symmetries $C$ and $P$, $H$ and $M$ mix only if
they share the same spin, $C$, and $P$ eigenvalues.  Therefore, the
observation of an invisible decay not only constrains the mass of the
dark matter and mediator, but may also uniquely identify the spin, $C$,
and $P$ of the mediator.  An invisible decay does not have sensitivity
to the spin of the Dark Matter itself.
%Related measurements where a visible object recoils against an
%invisible object such as $\Upsilon \rightarrow \gamma \chi \chi$ or $B
%\rightarrow K \chi \chi$ cannot determine the $J^{CP}$ of the mediator
%unless the intermediate decay into $\gamma M$ is 2-body, and therefore
%mediator mass $m_M^2 < m_H^2 - 2 m_H E_\gamma$.

With running B-factories BaBar and Belle having roughly 400
${\rm fb}^{-1}$ recorded, these experiments may already have tens of
thousands of DM production events, if the DM is kinematically accessible.

\section{Bottomonium production via ISR}
\label{sec:isr}
\begin{figure}[b]
    \includegraphics[scale=0.7]{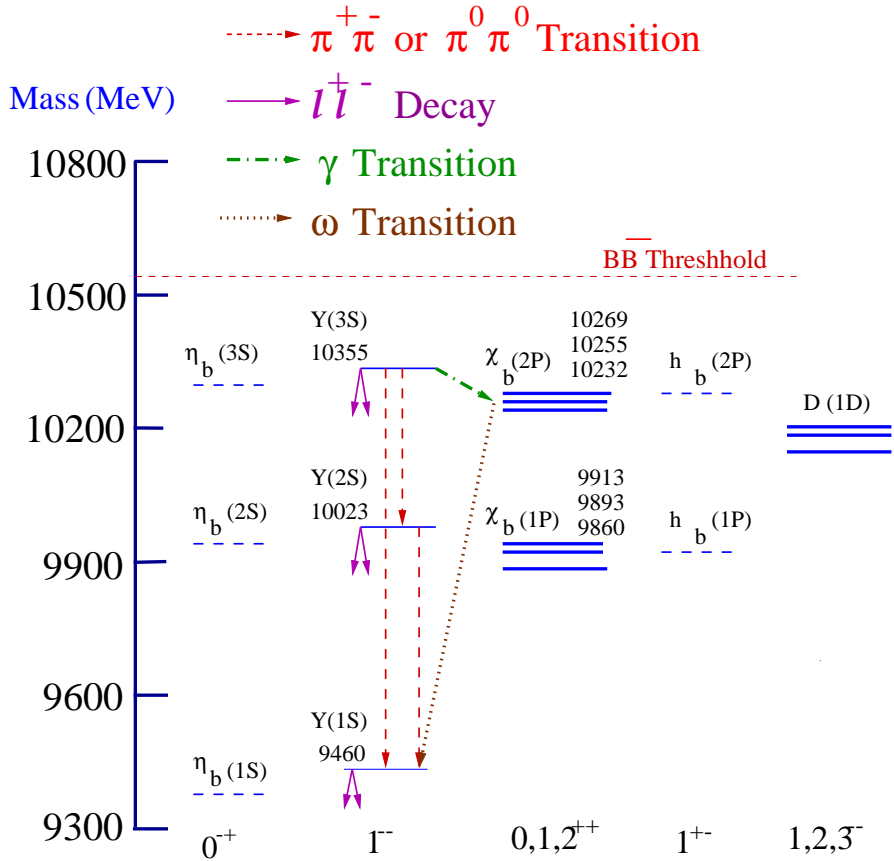}
    \caption{spectra for bottomonium,
    with the spin state on the horizontal axis}
    \label{fig:onia}
\end{figure}

Bottomonia can be identified by observing the particles emitted when it
makes radiative transitions to lighter bottomonia.  These transitions
are show in Fig.~\ref{fig:onia}.  Since the states are fairly narrow, the
energy of the photon radiated in a transition (or kinematics of a
particle pair) gives a clean way to select specific quarkonia
transitions.  The CLEO experiment has provided measurements of most of
the quarkonia transitions.~\cite{quarkonium}  The number of ISR
$\Upsilon$ production events collected by the BaBar and Belle
experiments is now competitive with that collected by the traditional
method of scanning the resonance.  Furthermore, off-peak data can be
used for the measurements we propose.

The ISR cross section for a particular final state $f$, with $e^+ e^-$
cross section $\sigma_f(s)$ is to first order~\cite{isrtheory}:
\begin{equation}
    \frac{d\sigma(s,x)}{dx} = W(s,x) \cdot \sigma_f(s(1-x))
\end{equation}
where $x= \frac{2 E_\gamma}{\sqrt{s}}$, $E_\gamma$ is the energy of the
ISR photon in the CM frame, and $\sqrt{s}$ is the CM
energy.  The function
\begin{equation}
\label{isrfrac}
    W(s,x)= \beta \left[(1+\delta) x^{(\beta-1)} - 1 + \frac{x}{2}\right]
\end{equation}
describes the energy spectrum of the ISR photons, where
$\beta=\frac{2\alpha}{\pi x}(2 \ln{\frac{\sqrt{s}}{m_e}}-1)$ and
$\delta$ take into account vertex and self-energy corrections; $\alpha$
is the electromagnetic coupling constant and $m_e$ is the mass of the
electron.  At the
$\Upsilon(4S)$ energy, $\beta=0.088$ and $\delta=0.067$.  By tagging the
ISR photon, B-factories can explore all of the vector $b \bar{b}$ bound
states.

This function (Eq.~\ref{isrfrac}) is strongly peaked in the forward and
backward directions, so ISR photons will be close to the beamline.
Assuming a detector acceptance of $-0.9 < \cos(\theta) < 0.9$, the
fraction of the nominal luminosity at the $\Upsilon(4S)$ resonance
delivered to the $\Upsilon(1S)$, $\Upsilon(2S)$ and $\Upsilon(3S)$
resonances is $1.9\times 10^{-5}$, $3.2\times 10^{-5}$ and $5.0\times
10^{-5}$ respectively.  This results in hundreds of thousands of events per
resonance with current recorded luminosities.

If one does not require that the ISR photon is identified because it is
not in the detector acceptance, the production of lower resonances is
larger by a factor $5$ to $7$.\cite{isrexpt}  The fraction of the nominal luminosity at the
$\Upsilon(4S)$ delivered to the $\Upsilon(1S)$, $\Upsilon(2S)$, and
$\Upsilon(3S)$ resonances is $8.5\times 10^{-5}$, $1.5\times 10^{-4}$,
and $2.3\times 10^{-4}$.

One can then tag quarkonium states by looking for a particular radiative
transition such as $\Upsilon(2S) \rightarrow \chi_{b0}(1P)\gamma$ or
$\Upsilon(3S) \rightarrow \Upsilon(1S) \pi^+ \pi^-$.  One can examine
the decay modes of the tagged state in a decay-mode independent manner.

\section{The kinematics of ISR production with radiative decays}

The irreducible physics background for these invisible decays coming
from a pair of neutrinos and pions is
extremely small due to the fact that weak cross sections are suppressed
by $(M_\Upsilon/M_W)^2 \simeq 0.01$, the final state has high
multiplicity, and our signal has resonant enhancement.
For example $\sigma(e^+ e^- \rightarrow \pi^+ \pi^- \nu
\bar{\nu}) \simeq 10^{-6}$ pb before applying any cuts.  Therefore, the
dominant backgrounds will come from unrelated processes that do {\it
not} actually have a neutrino or DM, or neutrinos from $\tau$ decays.  

The knowledge that resonances are formed in our signal gives us the
kinematic constraint that the square of the four-momenta forming a
resonance must be equal to the resonance mass-squared.  For the
production of a single $\Upsilon$ resonance via ISR and not observing
the ISR photon, there are 2 kinematic variables that are undetermined.
These variables are associated with the four-vectors of the ISR photon
and $\Upsilon$ which we presume decays invisibly.  With $n$ intermediate
resonances decaying radiatively to each other, 
in the event that the ISR photon is {\it unobserved}, we can
predict all but $2-n$ of these undetermined variables.  The most
important radiative decays of bottomonium are listed in 
Appendix~\ref{eventrates}, sorted by
cross section. Up to two intermediate resonances can be created with
sizeable event rates.

\subsection{Resonance Constraints}
\label{sec:resonances}

We can either observe the
ISR photon, or require that its angle with respect to the beamline is
consistent with it being outside the detector acceptance
using the kinematic constraints.

With one intermediate resonance and one radiative decay (e.g.
$\Upsilon(2S) \rightarrow \Upsilon(1S) \pi^+ \pi^-$), we define $M_1$
(e.g $M_{\Upsilon(2S)}$) to be the mass of the intermediate resonance
and $M_2$ (e.g. $M_{\Upsilon(1S)}$) to be the mass of the final
invisibly-decaying state.  We can predict all but one kinematic variable
using the measurement of the particles emitted in the radiative
transition and the beam constraint.  In the center-of-mass frame:
\begin{equation}
    \label{ethetaisreq}
    E_{\rm ISR} = \frac{s-M_1^2}{2 \sqrt{s}}, \qquad
    \cos \theta = \frac{\sqrt{s}}{p_{r}}
                 \frac{M_1^2-M_2^2+M_r^2}{s-M_1^2}
               - \frac{E_r}{p_r} 
                 \frac{s+M_1^2}{s-M_1^2}
\end{equation}
where $p_r^\mu = (E_r; \vec{p}_r)$ is the sum of the four momenta of all
the particles emitted in the radiative transition, $M_r^2 =
p_{r\mu}p_r^\mu$, $p_r = |\vec{p}_r|$, and $s=M^2_{\Upsilon(4S)}$ is the
center-of-mass energy.  Our predicted angle $\theta$ is the angle
between the ISR photon and $\vec{p}_r$.  $\cos \theta$ can be related
to $\Delta M^2 = M_1^2-M_2^2$.  However, when expressed as an angle it is
clear that it can still be used when the ISR photon is unobserved.

We can also invert Eq.~\ref{ethetaisreq} to come up with a cut on the
energy of the radiated system
\begin{equation}
    \label{eradbounds}
    \frac{M_1^2-M_2^2}{2 \sqrt{s}} < E_r <
    \frac{\sqrt{s}}{2}\left(1-\frac{M_2^2}{M_1^2}\right)
\end{equation}
which is useful for single photon transitions.  

Backgrounds without resonances can be distributed outside
the physical region $-1.0 \le \cos \theta \le 1.0$ since for background
without resonances, Eq.~\ref{ethetaisreq} does not describe any physical
angle at all.  Therefore, a cut requiring $-1.0 \le \cos \theta \le 1.0$
will suppress most backgrounds by a factor $10^2 - 10^3$ (c.f.
Table.~\ref{bgtable}).  If the ISR photon is unobserved we also
know that it must be outside the detector's acceptance.  Assuming the EM
calorimeter extends between $\theta_{\min}$ and $\theta_{\max}$ as seen
from the center-of-mass-frame, the angle between the ISR photon and the
beamline ($\theta_{\rm ISR}$) must satisfy $\theta_{\rm ISR} <
\theta_{min}$ or $\theta_{\rm ISR} > \theta_{max}$.  Since the angle
with respect to the beamline $\theta_r$ of $\vec{p}_r$ is measured, this
amounts to the restriction:
\begin{equation}
    \label{thetabounds}
    |\theta-\theta_r| < \theta_{min} 
    \qquad {\rm or} \qquad
    |\theta+\theta_r| > \theta_{max}
\end{equation}
Furthermore, the signal peaks in both these regions as $(\theta \pm
\theta_r)^2$.  This corresponds to the ISR photon being nearly parallel
to one of the beamlines.  

If the ISR photon is {\it observed} both $E_{\rm ISR}$ and
$\cos \theta$ are measured and can be directly compared to
(\ref{ethetaisreq}).  The final kinematic variable can be taken to be
the angle between the plane defined by the beamline and $\vec{p}_r$, and
the plane defined by beamline and $\vec{p}_{\rm ISR}$.  Only if the ISR
photon is observed can this angle be determined.  This angle will only
provide power in suppressing background if the background happens to
peak in this variable.  If the ISR photon is unobserved, this angle is
not knowable in principle.

When two intermediate resonances are formed, and two radiative decays
occur, all kinematic variables can be determined by measuring the energy
and momentum of the particles in the radiative decay.  The constraints
above still apply for the first transition.  The new
constraint available allows us to predict the angle between the ISR
photon and the {\it second} radiative decay:
\begin{equation}
    \label{thetaprime}
    \cos \theta^\prime = 
      \frac{\sqrt{s}}{|\vec{r}_2|} 
        \frac{M_2^2-M_3^2 + r_2^2 + 2 r_1 \cdot r_2}{s-M_1^2}
    - \frac{E_2}{|\vec{r}_2|} 
    \frac{s+M_1^2}{s-M_1^2}
\end{equation}

We can further apply the same trick as with $\cos \theta$ to require the
ISR photon to be in the beam line, replacing $\theta$ with
$\theta^\prime$ in Eq.~\ref{thetabounds}.

It should be noted that it is possible to emit two or more ISR photons.
In this case, Eqs.~\ref{ethetaisreq}-\ref{thetaprime} are accurate in
the limit that the ISR photons are collinear.

\subsection{Missing Momentum and Standard Model Decay Backgrounds}
Considering that there are invisible particles in the final states we
are interested in, there is an irreducible background when the final
state decays to visible Standard Model particles, but those particles
lie outside the detector acceptance.  A missing momentum cut can force
the transverse momentum of the final state to be large enough that one
can ensure that its decay products would be in the detector volume.
However, this is almost completely useless for machines running at the
$\Upsilon(4S)$ for the
following reason: assume the final state undergoes a 2-body decay,
those decay products lie exactly on the edge of a symmetric CLEO-like
detector at $\theta_{\min}$ and $\theta_{\max}=\pi-\theta_{\min}$, and
the final state is at rest at the center.  The cut required on the
transverse component of the sum of radiative transition particles is:
\begin{equation}
    (p_{r})_T > \frac{1}{2}(E_{CM}-E_{ISR}-E_r) \sin \theta_{min} \simeq
    2 {\rm GeV},
\end{equation}
which is larger than the visible energy ($E_{ISR}+E_r$) in any radiative
transition.  Here $E_{\rm CM}$ is the center of mass energy, $E_{\rm
ISR}$ is the energy of the ISR photon, and $E_r$ is the sum of energies
of particles emitted in the transition(s).  This background is further
discussed in Sec.~\ref{twobodybg}.

For a collider running at $\mathcal{O}(30 GeV)$, requiring the ISR
photon to be visible provides enough transverse momentum that the decay
products of the final state must lie in the detector acceptance.
However, the ISR production cross section of $\Upsilon$'s is reduced by
a factor $\sim 100$.  This requirement also eliminates most of the
two-photon background, as discussed in Sec.~\ref{photonfusionbg}.

\section{Invisible Upsilon Decays}

To demonstrate that invisible widths can be measured using ISR and
radiative decays, we concentrate on the modes $\Upsilon(nS) \rightarrow
\Upsilon(1S) \pi^+ \pi^-$ (n=2,3) for colliders running at the $\Upsilon(4S)$
since these modes have the largest
cross section (2.91pb, 0.784 pb).  Many decays\footnote{For instance 
$\Upsilon(3S) \rightarrow \Upsilon(2S) \gamma \gamma \rightarrow \Upsilon(1S) \gamma \gamma \pi^+ \pi^-$, 
$\Upsilon(3S) \rightarrow \Upsilon(2S) \rightarrow \pi^0 \pi^0 \rightarrow \Upsilon(1S) \pi^0 \pi^0 \pi^+ \pi^-$, 
$\Upsilon(3S) \rightarrow h_b(1P) \pi^+ \pi^- \rightarrow \eta_b(1S) \pi^+ \pi^- \gamma$, 
$\Upsilon(3S) \rightarrow \chi_{b0}(2P) \gamma \rightarrow \eta_b(1S) \gamma \eta$,
$\Upsilon(3S) \rightarrow \Upsilon(2S) \pi^+ \pi^- \rightarrow \chi_{b0}(1P) \pi^+ \pi^- \gamma$.
} 
have the visible topology $\pi^+
\pi^- + n\gamma$, $n\ge0$, which might be useful for triggering.  These
modes seem the most promising since the running B-factories BaBar and Belle
do not have the sensitivity in their calorimeter that CLEO had, and they
have excellent charged particle tracking resolution by design.  
These transitions were measured by CLEO.\cite{isrexpt}
A table
of possible transitions and their tagging signatures is presented in
Appendix~\ref{eventrates}.  The $\Upsilon(3S)$ mode has harder pions, and
therefore may have a higher reconstruction or triggering efficiency than
the $\Upsilon(2S)$ mode.

For background simulations we have used PYTHIA~\cite{pythia} and
CompHEP~\cite{comphep} with $\tau$ decays simulated with
TAUOLA~\cite{tauola}. We have smeared charged
tracks according to the BaBar detector resolution~\cite{babartdr}
\begin{equation}
    \frac{\sigma_{p_t}}{p_t} = 0.21\% \oplus 1.4\% p_t,
\end{equation}
charged tracks must have $p_t > 100$ MeV, and all objects must lie
within the detector $-0.87 < \cos \theta < 0.96$ from the center-of-mass
frame (i.e. BaBar geometry).  For photons we require $E > 20$ MeV and
smear their energy according to
\begin{equation}
    \frac{\sigma_E}{E} = 1.2\% \oplus \frac{1.0\%}{\left(E/{\rm GeV}\right)^{1/4}},
    \qquad
    \sigma_\theta = \sigma_\phi = 2 {\rm mr} \oplus \frac{3 {\rm mr}}{\sqrt{E/{\rm GeV}}}.
\end{equation}
Finally, for charged tracks we do not differentiate $\pi^+$, $e^+$,
$\mu^+$ or $K^+$ and assign each charged track to have a mass
$m_{\pi^+}$ after smearing its momentum, since tracking information is
reliable but particle ID is not.  These tracks generally are
soft enough that they do not enter the calorimeter, or enter at a
grazing angle.

The cuts proposed in Sec.~\ref{sec:resonances} are drastic and in
general can result in a background suppression of $10^5$ or more,
including effects of detector resolution.  
Due to the large size of these backgrounds, detailed detector resolution
effects and multiple scattering will be very important.  Therefore, each
background should be measured directly in order to estimate the signal
contamination.  Therefore, we present estimates of each background and
the level to which they can be suppressed including smearing.  The
exact numbers may change significantly when detector effects are taken
into account.

Aside from the purely kinematic constraints presented in
Sec.~\ref{sec:resonances},
there is no further angular information from the matrix element that is
useful to identify the signal.  In the case of a final state
$\Upsilon$ from a di-$\pi$ transition, there is no spin correlation
between the outgoing pions and the final $\Upsilon$.  The operator
involving a coupling between the polarization of the $\Upsilon$ and the
momenta of the pions is D-wave suppressed and measured to be very
small.~\cite{y3s1spipi} Furthermore, the $\pi \pi$ invariant mass
spectrum in the $\Upsilon(3S) \rightarrow \Upsilon(1S)\pi\pi$ transition
is not well understood theoretically and may involve another
intermediate state.~\cite{Guo:2004dt}  

Including all backgrounds and a realistic smearing of energy and
momenta, the measurements proposed can limit at $2 \sigma$ sensitivity
\begin{equation}
    BR(\Upsilon(1S) \rightarrow {\rm invisible}) < 0.113\%
\end{equation}
using the $\Upsilon(2S)$ mode and
\begin{equation}
    BR(\Upsilon(1S) \rightarrow {\rm invisible}) < 0.335\%
\end{equation}
using the $\Upsilon(3S)$ mode.  The combined $2 \sigma$ sensitivity is
then
\begin{equation}
    BR(\Upsilon(1S) \rightarrow {\rm invisible}) < 0.107\%.
\end{equation}

\begin{table}[t]
\begin{tabular}[t]{c||c||c||c||c||c}
    \multicolumn{6}{c}{Backgrounds to $\Upsilon(3S) \rightarrow \Upsilon(1S) \pi^+ \pi^-$} \\
    \hline
   cut & $\tau^+ \tau^-$ & %929 pb  & 
           $\gamma \gamma \rightarrow l^+ l^-$ &
           $\gamma \gamma \rightarrow {\rm hadrons}$ &
           $\gamma \gamma \rightarrow l^+ l^- \gamma$ &
           $\Upsilon(1S) \rightarrow l^+ l^-$ \\
\hline
$\pi^+ \pi^- \gamma$ selection &  71.8 pb  & 228 fb  & 866 fb  & 44.9 pb  & 1.41 fb \\
$-1.1 < \cos \theta < 1.1$     &  120.8 fb  & $< 0.1$ fb  & 3.7 fb  & 1.40 pb  & 1.39 fb \\
$|\cos \theta - \cos \theta_{\rm meas}| < 0.15$  & 16.2 fb  & $< 0.1$ fb  & 0.5 fb  & 197 fb  & 1.20 fb \\
$|E_\gamma-E_{\rm ISR}| < 6 {\rm MeV}$  & $< 0.1$ fb  & $< 0.1$ fb  & $< 0.1$ fb  & 5.4 fb  & 1.06 fb\\
\hline 
    \multicolumn{6}{c}{Backgrounds to $\Upsilon(2S) \rightarrow \Upsilon(1S) \pi^+ \pi^-$} \\
\hline
$\pi^+ \pi^- \gamma$ selection & 71.8 pb  & 228 fb  & 866 fb  & 44.9 pb  & 3.97 fb\\
$-1.1 < \cos \theta < 1.1$ & 60.7 fb  & $ < 0.1$ fb  & 0.8 fb  & 3.65 pb  & 3.97 fb\\
$|\cos \theta - \cos \theta_{\rm meas}| < 0.035$ & 2.1 fb  & $< 0.1$ fb  & $< 0.1$ fb  & 108 fb  & 2.80 fb\\
$|E_\gamma-E_{\rm ISR}| < 15 {\rm MeV}$ &  $< 0.1$ fb  & $< 0.1$ fb  & $< 0.1$ fb  & 1.6 fb  & 2.52 fb\\
\end{tabular}
\caption{Dominant backgrounds to the processes $e^+ e^- \rightarrow
\Upsilon(nS) \gamma_{ISR} \rightarrow \Upsilon(1S) \pi^+ \pi^-
\gamma_{ISR}$.  Cuts are shown in the left column, followed by the effective
remaining cross section in each background channel after the cut.  Each line
includes all the cuts above it.  The photon is assumed visible, and any charged
track is assumed to be a pion.  For the $\Upsilon(2S)$ mode, the 2$\sigma$
sensitivity on BR($\Upsilon(1S) \rightarrow invisible) \simeq 0.113\%$; for the
$\Upsilon(3S)$ mode, the 2$\sigma$ sensitivity is $\simeq 0.335\%$.}
\label{bgtable} 
\end{table}

The dominant backgrounds are discussed in the following subsections.
Their numeric importance and cuts needed to suppress them is summarized
in Table~\ref{bgtable}.

\subsection{Photon Fusion Background}
\label{photonfusionbg}

The two photon fusion process occurs when both incoming beams emit a
photon and those photons annihilate into electrons, muons, taus, or hadrons.
This cross section is very large, in the hundreds of nanobarns.
Furthermore, our signal spans the region $0 < Q^2 < 1$ GeV$^2$ in which
non-perturbative QCD effects dominate hadron production.  Due to this, a
reliable simulation of hadron production is not possible and in any
case should not be relied upon due to non-perturbative effects.  This
background must be measured directly.

To demonstrate that this background can be overcome, we simulate 10
fb$^{-1}$ of
the lepton\footnote{Here lepton refers to an $e$ or $\mu$.} production
processes $e^+ e^- \rightarrow e^+ e^- l^+ l^-$, $e^+ e^- \rightarrow
e^+ e^- + {\rm hadrons}$ and $e^+ e^- \rightarrow
e^+ e^- l^+ l^- \gamma$.  We simulate the first two backgrounds using PYTHIA,
and the second using CompHEP~\footnote{The $\gamma
\gamma \rightarrow l^+ l^- \gamma$ process is not included in
PYTHIA.} in the equivalent photon approximation.~\cite{Budnev:1974de} At
these low energies, $\pi/\mu$ separation is generally unreliable since
the muon is not energetic enough to reach the outer muon detector.  The
$l^+ l^-$ cross section is also about an order of magnitude larger than
the $\pi^+ \pi^-$ cross section, making the $l^+ l^-$ the most important
background in any case.  The $\mu^+ \mu^-$ cross section is 38.5 nb for
the BaBar detector geometry, assuming both muons are visible in the
detector volume.  This is sufficiently large that it overwhelms the
physics signal that is normally triggered on at the B-factories (about 1
nb).  Therefore, the rate must be reduced at the trigger level.
Requiring an extra visible photon reduces this cross section to a
triggerable level.  In the case of the $e^+ e^- \rightarrow e^+ e^- l^+
l^-$ background, an extra photon comes from Initial/Final State
Radiation.

Di-lepton events produced in photon fusion have the characteristic that
the leptons are back-to-back in the plane perpendicular to the beamline.
By contrast, our signal is a 3-body decay, so only a small fraction are
back-to-back.  Therefore, $\vec{p}_r=0$ and $\cos \theta$ (c.f.
Eq.~\ref{ethetaisreq}) will be very large.  As can be seen in
Table~\ref{bgtable} a cut on $\cos \theta$ alone can remove this
background.  If detector resolution effects cause this background to
bleed into the signal region, a cut $\Delta \phi_{ll} <
\pi$, where $\phi$ is the angle between the leptons in the plane
perpendicular to the beamline can also remove this background.
Therefore, this background should be carefully studied.  This can be done
by identifying singly-tagged photon fusion events, where one of the
initial electrons is deflected into the detector.  Experiments such as
DELPHI have also employed far-forward particle detectors to identify the
photon fusion signal when the electrons are deflected by a small angle
(known as the Small Angle Tagger and Very Small Angle Tagger).

\subsection{di-$\tau$ Background}

The only irreducible physics background to these processes that have
true neutrinos comes from $\tau$ decays.  The dominant source is $e^+
e^- \rightarrow \tau^+ \tau^-$ via a virtual photon, where both $\tau$'s
decay to pions.  The total $\tau^+ \tau^-$ cross section is 993 pb at
tree level.  $\tau$'s can also be produced in photon fusion, with a
cross-section of approximately 22 pb.

For example, we consider the di-$\tau$ background to the process
$\Upsilon(3S) \rightarrow \Upsilon(1S) \pi^+ \pi^-$.  We require exactly
two charged tracks and one photon visible in the detector.  Generally
the photon comes from a $\pi^0$ decay in which the other photon is
outside the detector's acceptance, or final/initial state radiation.

The effect of the kinematic cuts proposed in Sec.~\ref{sec:resonances}
are shown in Table \ref{bgtable}.  The di-$\tau$ background generally
has very different kinematics than our signal, as well as extra
$\pi^0$'s.  It can be reduced below 0.1 fb with these cuts.

\subsection{Two-Body Decay Background}
\label{twobodybg}
A background to all processes is true resonance production where the
final state resonance decays via any 2-body decay and its decay
products lie outside the detector acceptance.  This background is
irreducible, but is accurately measured in events with both the
radiative decay of interest and the final hadron decaying to a 2-body
state.  This gives roughly 10 times the statistics on measuring this
background, so it can be subtracted.

For 2-body
decays this amounts to an irreducible background that has a 
branching ratio $f_2 \Omega$.  Here $f_2$ is the fraction of 2-body final-state
bottomonium decays plus multi-body decays which are arranged
such that all decay products are outside the detector
acceptance.
$\Omega$ is the fraction of the solid angle covered by the detector.
This background can be directly measured by relying on the
sample of non-invisibly decaying final state particles provided by the
ISR + radiative transition technique.  We take $f_2=5\%$ and
$\Omega=91.5\%$ for the BaBar detector geometry.
The final decay is uncorrelated
to its production mechanism and the radiative transition, and therefore is
isotropic in detector angle $\theta$ and $\phi$.  Events with a
radiative transition and visible bottomonium
decay give a measurement of {\it all} decay channels of the final state
particle, including effects of detector resolution for the radiative
transition, ISR smear, and multi-body decays.  
It should be
noted that one cannot simply take this background sample and pretend the
beamline bisects the detector in a different direction.  The asymmetric
boost of modern B-factories changes the size and area of the would-be
beamline and this must be taken into account.

\subsection{Drell-Yan}

Direct production of $\nu \bar{\nu}$ is small since the neutrinos must
come from a $Z$ or $W^\pm$, which are heavy.  For instance, BR$(\Upsilon(1S)
\rightarrow \nu \bar{\nu}) \simeq 1 \times 10^{-5}$.~\cite{sminvisible}
Only the vector resonances have a sizeable branching fraction to neutrinos,
since they can mix directly with the $Z$.  Scalars and pseudo-scalars can only
emit neutrinos in loop suppressed processes.

Modern B-factories do not have the sensitivity to test this branching ratio,
and therefore it is not a background.

\section{Conclusions}

Measurements of invisible branching ratios of mesons are extremely
important, given the established evidence for Dark Matter (DM) and the
knowledge that most DM scenarios require some Standard Model-DM
interaction.  Given tight constraints on flavor changing neutral
currents, the most important mesons to examine are flavor neutral bound
states of quarks.  Running high-luminosity B-factories motivate
looking for invisible decays of bottomonium first.

ISR and radiative decays provide a powerful method to measure the
invisible branching fractions of the bottomonium resonances.  If the
DM is lighter than $M_\Upsilon/2$, annihilation of DM
into standard model particles is expected to have a picobarn-scale cross
section.  While the sensitivity achievable is not capable of measuring
the Standard Model $\Upsilon \rightarrow \nu \bar{\nu}$, decays to dark
matter should be significantly stronger than decays to neutrinos, due to
the $(M_\Upsilon/M_Z)^2$ suppression of the Standard Model process.
These techniques can limit $BR(\Upsilon(1S) \rightarrow {\rm invisible})
\lsim 0.1\%$, which is sensitive enough to discover dark matter if it
couples in this manner.

DM with a mass $M_\chi < 5$ GeV is generally allowed in
models.  Direct detection experiments are very
insensitive in this mass region, and would also be insensitive if dark
matter preferentially couples to heavy quarks.  Therefore, alternative methods to
discover DM are required if the DM is this light.  We
strongly encourage experimental teams at BaBar, Belle, and CLEO to
pursue these techniques.

\vskip 0.5in
\vbox{
\noindent{ {\bf Acknowledgments} } \\
\noindent
We thank Jack Gunion, Tao Han, Dan Hooper, and Steve Sekula for useful
discussions; Dave Mattingly and Steve Sekula carefully reading this
manuscript.  This work was supported in part by DOE grant
DE--FG03--91ER--40674, the Davis Institute for High Energy Physics, and
the U.C. Davis Dean's office.}

\pagebreak
\appendix
\renewcommand{\thesubsection}{A-\arabic{subsection}}
\section{Bottomonium Event Rates}
\label{eventrates}
In the following we give the expected cross section for bottomonium
production assuming $E_{\rm CM}=M_{\Upsilon(4S)}=10.58$ GeV.  It should
be noted that both on-peak and off-peak data can be used for this
analysis.

For each final state, a ``tagging topology'' is given, which is the set
of particles visible in the detector's acceptance.  In each case the particles
in the tagging topology have a well-defined kinematics as outlined in
Sec.~\ref{sec:resonances}.  In cases where $\gamma_{\rm ISR}$ is not listed
in the tagging topology, the cross section includes both when the ISR
photon is visible, and when it lies outside the detector acceptance.
When $\gamma_{\rm ISR}$ is listed, the cross section corresponds to
requiring the ISR photon to be visible for the BaBar detector geometry,
$-0.87 < \cos \theta < 0.96$ in the center-of-mass frame.

We catalog only existing, measured resonances and transitions of
$b\bar{b}$ quarkonia here (with the exception of the undiscovered
$\eta_b$).  Other decay chains will be possible involving radiative
decays of the $\Upsilon(4S)$ when those decays are discovered.  This
clean method of tagging the initial state will allow the discovery and
cataloging of many more radiative decays, improving statistics from what
is listed below.

\subsection{Vector Mediated Dark Matter}
\begin{tabular}{c|c|cc|r}
    & & & & \\
    Final state    & Decay chain & \multicolumn{2}{c|}{Tagging topology}  &
    $\sigma_\Upsilon$(pb) \\
    \hline
$\Upsilon(1S)$&&$\gamma_{\rm ISR}$ & & 3.034\\
& $\Upsilon(2S) \rightarrow \Upsilon(1S)$ & $\pi^+\pi^-$ & $$ & 2.91\\
& $\Upsilon(2S) \rightarrow \Upsilon(1S)$ & $\pi^0\pi^0$ & $$ & 1.4\\
& $\Upsilon(3S) \rightarrow \Upsilon(1S)$ & $\pi^+\pi^-$ & $$ & 0.784\\
& $\Upsilon(2S) \rightarrow \chi_{b1}(1P) \rightarrow \Upsilon(1S)$ & $\gamma$ & $\gamma$ & 0.369\\
& $\Upsilon(3S) \rightarrow \Upsilon(1S)$ & $\pi^0\pi^0$ & $$ & 0.36\\
& $\Upsilon(2S) \rightarrow \chi_{b2}(1P) \rightarrow \Upsilon(1S)$ & $\gamma$ & $\gamma$ & 0.239\\
& $\Upsilon(3S) \rightarrow \chi_{b1}(2P) \rightarrow \Upsilon(1S)$ & $\gamma$ & $\gamma$ & 0.168\\
& $\Upsilon(3S) \rightarrow \Upsilon(2S) \rightarrow \Upsilon(1S)$ & $\gamma\gamma$ & $\pi^+\pi^-$ & 0.165\\
& $\Upsilon(3S) \rightarrow \chi_{b2}(2P) \rightarrow \Upsilon(1S)$ & $\gamma$ & $\gamma$ & 0.142\\
& $\Upsilon(3S) \rightarrow \Upsilon(2S) \rightarrow \Upsilon(1S)$ & $\pi^+\pi^-$ & $\pi^+\pi^-$ & 0.0921\\
& $\Upsilon(3S) \rightarrow \Upsilon(2S) \rightarrow \Upsilon(1S)$ & $\gamma\gamma$ & $\pi^0\pi^0$ & 0.0788\\
& $\Upsilon(3S) \rightarrow \Upsilon(2S) \rightarrow \Upsilon(1S)$ & $\pi^0\pi^0$ & $\pi^+\pi^-$ & 0.0658\\
& $\Upsilon(3S) \rightarrow \Upsilon(2S) \rightarrow \Upsilon(1S)$ & $\pi^+\pi^-$ & $\pi^0\pi^0$ & 0.0441\\
& $\Upsilon(3S) \rightarrow \chi_{b1}(2P) \rightarrow \Upsilon(1S)$ & $\gamma$ & $\omega$ & 0.0322\\
& $\Upsilon(3S) \rightarrow \Upsilon(2S) \rightarrow \Upsilon(1S)$ & $\pi^0\pi^0$ & $\pi^0\pi^0$ & 0.0315\\
& $\Upsilon(3S) \rightarrow \chi_{b2}(2P) \rightarrow \Upsilon(1S)$ & $\gamma$ & $\omega$ & 0.0219\\
& $\Upsilon(3S) \rightarrow \chi_{b0}(2P) \rightarrow \Upsilon(1S)$ & $\gamma$ & $\gamma$ & 0.0107\\
\hline Total & \multicolumn{3}{l|}{} & 6.912\\\hline$\Upsilon(2S)$&&$\gamma_{\rm ISR}$ &  &2.465\\
& $\Upsilon(3S) \rightarrow \Upsilon(2S)$ & $\gamma\gamma$ & $$ & 0.875\\
& $\Upsilon(3S) \rightarrow \Upsilon(2S)$ & $\pi^+\pi^-$ & $$ & 0.49\\
& $\Upsilon(3S) \rightarrow \chi_{b1}(2P) \rightarrow \Upsilon(2S)$ & $\gamma$ & $\gamma$ & 0.415\\
& $\Upsilon(3S) \rightarrow \Upsilon(2S)$ & $\pi^0\pi^0$ & $$ & 0.35\\
& $\Upsilon(3S) \rightarrow \chi_{b2}(2P) \rightarrow \Upsilon(2S)$ & $\gamma$ & $\gamma$ & 0.323\\
& $\Upsilon(3S) \rightarrow \chi_{b0}(2P) \rightarrow \Upsilon(2S)$ & $\gamma$ & $\gamma$ & 0.0545\\
\hline Total & \multicolumn{3}{l|}{} & 2.508\\\hline
\end{tabular}
\subsection{Pseudoscalar Mediated Dark Matter}
\begin{tabular}{c|c|cc|r|c}
    & & & & \\
    Final state    & Decay chain & \multicolumn{2}{l|}{Tagging topology}  &
    $\sigma_\Upsilon$(pb) \\
    \hline
$\eta_b(1S)$
& $\Upsilon(3S) \rightarrow h_b(1P) \rightarrow \eta_b(1S)$ & $\pi^+\pi^-$ & $\gamma$ & 0.00874\\
& $\Upsilon(3S) \rightarrow h_b(1P) \rightarrow \eta_b(1S)$ & $\pi^0$ & $\gamma$ & 0.00236\\
& $\Upsilon(3S) \rightarrow \chi_{b0}(2P) \rightarrow \eta_b(1S)$ & $\gamma$ & $\eta$ & 0.00213\\
\hline Total & \multicolumn{3}{l|}{} & 0.013\\\hline
\end{tabular}
\subsection{Scalar Mediated Dark Matter}
\begin{tabular}{c|c|cc|r|c}
    & & & & \\
    Final state    & Decay chain & \multicolumn{2}{l|}{Tagging topology}  &
    $\sigma_\Upsilon$(pb) \\
    \hline
$\chi_{b0}(1P)$
& $\Upsilon(2S) \rightarrow \chi_{b0}(1P)$ & $\gamma_{\rm ISR}$ & $\gamma$ & 0.0937\\
& $\Upsilon(3S) \rightarrow \Upsilon(2S) \rightarrow \chi_{b0}(1P)$ & $\gamma\gamma$ & $\gamma$ & 0.0333\\
& $\Upsilon(3S) \rightarrow \Upsilon(2S) \rightarrow \chi_{b0}(1P)$ & $\pi^+\pi^-$ & $\gamma$ & 0.0186\\
& $\Upsilon(3S) \rightarrow \Upsilon(2S) \rightarrow \chi_{b0}(1P)$ & $\pi^0\pi^0$ & $\gamma$ & 0.0133\\
\hline Total & \multicolumn{3}{l|}{} & 0.159\\\hline$\chi_{b0}(2P)$
& $\Upsilon(3S) \rightarrow \chi_{b0}(2P)$ & $\gamma_{\rm ISR}$ & $\gamma$ & 0.188\\
\hline Total & \multicolumn{3}{l|}{} & 0.188\\\hline
\end{tabular}


\begin{thebibliography}{99}

%\cite{Bertone:2004pz}
\bibitem{dmreview}
G.~Bertone, D.~Hooper and J.~Silk,
%``Particle dark matter: Evidence, candidates and constraints,''
Phys.\ Rept.\  {\bf 405}, 279 (2005)
[arXiv:hep-ph/0404175].
%%CITATION = HEP-PH 0404175;%%

\bibitem{pdg}
%\cite{Eidelman:2004wy}
%\bibitem{Eidelman:2004wy}
S.~Eidelman {\it et al.}  [Particle Data Group Collaboration],
%``Review of particle physics,''
Phys.\ Lett.\ B {\bf 592}, 1 (2004).
%%CITATION = PHLTA,B592,1;%%

%\cite{Jean:2003ci}
\bibitem{Jean:2003ci}
P.~Jean {\it et al.},
%``Early SPI/INTEGRAL measurements of galactic 511 keV line emission from
%positron annihilation,''
Astron.\ Astrophys.\  {\bf 407}, L55 (2003)
[arXiv:astro-ph/0309484].
%%CITATION = ASTRO-PH 0309484;%%

%\cite{Strong:2004de}
\bibitem{Strong:2004de}
  A.~W.~Strong, I.~V.~Moskalenko and O.~Reimer,
  %``Diffuse Galactic continuum gamma rays. A model compatible with EGRET data
  %and cosmic-ray measurements,''
  Astrophys.\ J.\  {\bf 613}, 962 (2004)
  [arXiv:astro-ph/0406254].
  %%CITATION = ASTRO-PH 0406254;%%

%\cite{Beacom:2004pe}
\bibitem{Beacom:2004pe}
  J.~F.~Beacom, N.~F.~Bell and G.~Bertone,
  %``Gamma-ray constraint on Galactic positron production by MeV dark matter,''
  arXiv:astro-ph/0409403.
  %%CITATION = ASTRO-PH 0409403;%%

%\cite{Gelmini:2004gm}
\bibitem{dmflows}
G.~Gelmini and P.~Gondolo,
%``DAMA dark matter detection compatible with other searches,''
arXiv:hep-ph/0405278.
%%CITATION = HEP-PH 0405278;%%

%\cite{Bottino:2002ry}
\bibitem{Bottino:2002ry}
A.~Bottino, N.~Fornengo and S.~Scopel,
%``Light relic neutralinos,''
Phys.\ Rev.\ D {\bf 67}, 063519 (2003)
[arXiv:hep-ph/0212379];\\
%%CITATION = HEP-PH 0212379;%%
%\cite{Bottino:2003iu}
%\bibitem{Bottino:2003iu}
A.~Bottino, F.~Donato, N.~Fornengo and S.~Scopel,
%``Lower bound on the neutralino mass from new data on CMB and  implications for
%relic neutralinos,''
Phys.\ Rev.\ D {\bf 68}, 043506 (2003)
[arXiv:hep-ph/0304080].
%%CITATION = HEP-PH 0304080;%%

\bibitem{ups1stogammainvis}
%\cite{Balest:1994ch}
%\bibitem{Balest:1994ch}
R.~Balest {\it et al.}  [CLEO Collaboration],
%``Upsilon (1s) $\to$ gamma + noninteracting particles,''
Phys.\ Rev.\ D {\bf 51}, 2053 (1995).
%%CITATION = PHRVA,D51,2053;%%

\bibitem{mySinglinoDM}
%\cite{Gunion:2005rw}
%\bibitem{Gunion:2005rw}
  J.~F.~Gunion, D.~Hooper and B.~McElrath,
  %``Light neutralino dark matter in the NMSSM,''
  arXiv:hep-ph/0509024.
  %%CITATION = HEP-PH 0509024;%%

\bibitem{invisiblehiggs}
%\cite{unknown:2001xz}
%\bibitem{unknown:2001xz}
  [LEP Higgs Working for Higgs boson searches Collaboration],
%``Searches for invisible Higgs bosons: Preliminary combined results using  LEP
%data collected at energies up to 209-GeV,''
arXiv:hep-ex/0107032.
%%CITATION = HEP-EX 0107032;%%

\bibitem{lepsinglephoton}
%\cite{Buskulic:1993ke}
%\bibitem{Buskulic:1993ke}
D.~Buskulic {\it et al.}  [ALEPH Collaboration],
%``A Direct measurement of the invisible width of the Z from single photon
%counting,''
Phys.\ Lett.\ B {\bf 313}, 520 (1993).
%%CITATION = PHLTA,B313,520;%%

\bibitem{nmssm} 
J. Ellis, J.F. Gunion, H.E. Haber, L. Roszkowski, and F. Zwirner
Phys. Rev. {\bf D39}, 844 (1989);
H.P. Nilles, M. Srednicki and D. Wyler, Phys. Lett. {\bf 120 B}
(1983) 346; \\
M. Drees, Int. J. Mod. Phys. {\bf A 4} (1989) 3635;\\ 
U. Ellwanger and M. Rausch de Traubenberg, Z. Phys. {\bf C 53} (1992)
521;\\ 
P.N. Pandita,
Z. Phys. {\bf C 59} (1993) 575; Phys. Lett. {\bf B 318} (1993) 338; \\
T. Elliott, S.F. King and P.L. White, Phys. Rev. {\bf D 49} (1994)
2435;\\
%\cite{Ellwanger:1997jj}
%\bibitem{Ellwanger:1997jj}
U.~Ellwanger and C.~Hugonie,
%``Neutralino cascades in the (M+1)SSM,''
Eur.\ Phys.\ J.\ C {\bf 5}, 723 (1998)
[arXiv:hep-ph/9712300]; \\
%%CITATION = HEP-PH 9712300;%%
%\cite{Panagiotakopoulos:2000wp}
%\bibitem{Panagiotakopoulos:2000wp}
C.~Panagiotakopoulos and A.~Pilaftsis,
%``Higgs scalars in the minimal non-minimal supersymmetric standard model,''
Phys.\ Rev.\ D {\bf 63}, 055003 (2001)
[arXiv:hep-ph/0008268];\\
%%CITATION = HEP-PH 0008268;%%
%\cite{Dedes:2000jp}
%\bibitem{Dedes:2000jp}
A.~Dedes, C.~Hugonie, S.~Moretti and K.~Tamvakis,
%``Phenomenology of a new minimal supersymmetric extension of the
%standard
%model,''
Phys.\ Rev.\ D {\bf 63}, 055009 (2001)
[arXiv:hep-ph/0009125].
%%CITATION = HEP-PH 0009125;%%

\bibitem{Boehm:2003hm}
%\cite{Boehm:2003hm}
C.~Boehm and P.~Fayet,
%``Scalar dark matter candidates,''
Nucl.\ Phys.\ B {\bf 683}, 219 (2004)
[arXiv:hep-ph/0305261].
%%CITATION = HEP-PH 0305261;%%

\bibitem{morelightdm}
%\cite{Boehm:2003bt}
%\bibitem{Boehm:2003bt}
C.~Boehm, D.~Hooper, J.~Silk, M.~Casse and J.~Paul,
%``MeV dark matter: Has it been detected?,''
Phys.\ Rev.\ Lett.\  {\bf 92}, 101301 (2004)
[arXiv:astro-ph/0309686];\\
%%CITATION = ASTRO-PH 0309686;%%
Celine Boehm, Yago Ascasibar,
%``More evidence in favour of Light Dark Matter particles?,''
[arXiv:hep-ph/0408213].

%\cite{Fayet:1991ux}
\bibitem{Fayet:1991ux}
P.~Fayet and J.~Kaplan,
%``Can the upsilon decay into cosmions?,''
Phys.\ Lett.\ B {\bf 269}, 213 (1991).
%%CITATION = PHLTA,B269,213;%%

\bibitem{isrexpt}
%\cite{Benayoun:1999hm}
%\bibitem{Benayoun:1999hm}
M.~Benayoun, S.~I.~Eidelman, V.~N.~Ivanchenko and Z.~K.~Silagadze,
%``Spectroscopy at B-factories using hard photon emission,''
Mod.\ Phys.\ Lett.\ A {\bf 14}, 2605 (1999)
[arXiv:hep-ph/9910523];\\
%%CITATION = HEP-PH 9910523;%%
%\cite{Kuhn:2001nn}
%\bibitem{Kuhn:2001nn}
J.~H.~Kuhn,
%``Measuring sigma(e+ e- $\to$ hadrons) with tagged photons at  electron
%positron colliders,''
Nucl.\ Phys.\ Proc.\ Suppl.\  {\bf 98}, 289 (2001)
[arXiv:hep-ph/0101100];\\
%%CITATION = HEP-PH 0101100;%%
%\cite{Solodov:2002xu}
%\bibitem{Solodov:2002xu}
E.~P.~Solodov  [BABAR collaboration],
%``Study of e+ e- collisions in the 1.5-GeV - 3-GeV cm energy region using  ISR
%at BaBar,''
in {\it Proc. of the $e^+ e^-$ Physics at Intermediate Energies Conference } ed. Diego Bettoni,
eConf {\bf C010430}, T03 (2001)
[arXiv:hep-ex/0107027].

\bibitem{isrtheory}
%\cite{Baier:1973ms}
%\bibitem{Baier:1973ms}
  V.~N.~Baier, V.~S.~Fadin and V.~A.~Khoze,
  %``Quasireal Electron Method In High-Energy Quantum Electrodynamics,''
  Nucl.\ Phys.\ B {\bf 65}, 381 (1973);\\
  %%CITATION = NUPHA,B65,381;%%
%\cite{Bonneau:1971mk}
%\bibitem{Bonneau:1971mk}
  G.~Bonneau and F.~Martin,
  %``Hard Photon Emission In E+ E- Reactions,''
  Nucl.\ Phys.\ B {\bf 27}, 381 (1971);\\
  %%CITATION = NUPHA,B27,381;%%
%%CITATION = HEP-EX 0107027;%%
%\cite{Rodrigo:2001jr}
%\bibitem{Rodrigo:2001jr}
G.~Rodrigo, A.~Gehrmann-De Ridder, M.~Guilleaume and J.~H.~Kuhn,
%``NLO QED corrections to ISR in e+ e- annihilation and the measurement of
%sigma(e+ e- $\to$ hadrons) using tagged photons,''
Eur.\ Phys.\ J.\ C {\bf 22}, 81 (2001)
[arXiv:hep-ph/0106132].
%%CITATION = HEP-PH 0106132;%%

\bibitem{bsme}
%\cite{Bird:2004ts}
%\bibitem{Bird:2004ts}
  C.~Bird, P.~Jackson, R.~Kowalewski and M.~Pospelov,
  %``Search for dark matter in b $\to$ s transitions with missing energy,''
  Phys.\ Rev.\ Lett.\  {\bf 93}, 201803 (2004)
  [arXiv:hep-ph/0401195].
  %%CITATION = HEP-PH 0401195;%%

\bibitem{sminvisible}
%\cite{Chang:1997tq}
%\bibitem{Chang:1997tq}
L.~N.~Chang, O.~Lebedev and J.~N.~Ng,
%``On the invisible decays of the Upsilon and J/psi resonances,''
Phys.\ Lett.\ B {\bf 441}, 419 (1998)
[arXiv:hep-ph/9806487].
%%CITATION = HEP-PH 9806487;%%

%\cite{Berger:2000mp}
\bibitem{Berger:2000mp}
  E.~L.~Berger, B.~W.~Harris, D.~E.~Kaplan, Z.~Sullivan, T.~M.~P.~Tait and C.~E.~M.~Wagner,
  %``Low energy supersymmetry and the Tevatron bottom-quark cross section,''
  Phys.\ Rev.\ Lett.\  {\bf 86}, 4231 (2001)
  [arXiv:hep-ph/0012001].
  %%CITATION = HEP-PH 0012001;%%
%\cite{Berger:2000mp}

%\cite{Janot:2004cy}
\bibitem{Janot:2004cy}
  P.~Janot,
  %``The light sbottom mass window revisited,''
  arXiv:hep-ph/0403157.
  %%CITATION = HEP-PH 0403157;%%

%\cite{Gogoladze:2002xp}
\bibitem{Gogoladze:2002xp}
I.~Gogoladze, J~Lykken, C.~Macesanu and S.~Nandi,
%``Implications of a massless neutralino for neutrino physics,''
Phys.\ Rev.\ D {\bf 68}, 073004 (2003)
[arXiv:hep-ph/0211391].
%%CITATION = HEP-PH 0211391;%%

%\cite{Lee:1977ua}
\bibitem{Lee:1977ua}
B.~W.~Lee and S.~Weinberg,
%``Cosmological Lower Bound On Heavy-Neutrino Masses,''
Phys.\ Rev.\ Lett.\  {\bf 39}, 165 (1977).
%%CITATION = PRLTA,39,165;%%

\bibitem{wmap}
%\cite{Bennett:2003bz}
%\bibitem{Bennett:2003bz}
  C.~L.~Bennett {\it et al.},
  %``First Year Wilkinson Microwave Anisotropy Probe (WMAP) Observations:
  %Preliminary Maps and Basic Results,''
  Astrophys.\ J.\ Suppl.\  {\bf 148}, 1 (2003)
  [arXiv:astro-ph/0302207].
  %%CITATION = ASTRO-PH 0302207;%%
%\cite{Spergel:2003cb}
%\bibitem{Spergel:2003cb}
  D.~N.~Spergel {\it et al.}  [WMAP Collaboration],
  %``First Year Wilkinson Microwave Anisotropy Probe (WMAP) Observations:
  %Determination of Cosmological Parameters,''
  Astrophys.\ J.\ Suppl.\  {\bf 148}, 175 (2003)
  [arXiv:astro-ph/0302209].
  %%CITATION = ASTRO-PH 0302209;%%

\bibitem{quarkonium}
%\cite{Galik:2004mi}
%\bibitem{Galik:2004mi}
R.~S.~Galik,
%``Quarkonium production and decay,''
arXiv:hep-ph/0408190;\\
%%CITATION = HEP-PH 0408190;%%
%\cite{Duboscq:2004uj}
%\bibitem{Duboscq:2004uj}
J.~E.~Duboscq  [CLEO Collaboration],
%``CLEO results on upsilons,''
arXiv:hep-ex/0405033;\\
%%CITATION = HEP-EX 0405033;%%
%\cite{Cronin-Hennessy:2003wj}
%\bibitem{Cronin-Hennessy:2003wj}
D.~Cronin-Hennessy {\it et al.}  [CLEO Collaboration],
%``Observation of the hadronic transitions chi/b1,2(2P) $\to$ omega
%Upsilon(1S),''
arXiv:hep-ex/0311043; \\
%%CITATION = HEP-EX 0311043;%%
%\cite{Severini:2003qw}
%\bibitem{Severini:2003qw}
H.~Severini {\it et al.}  [CLEO Collaboration],
%``Observation of the hadronic transitions chi/b(1,2)(2P) $\to$ omega
%Upsilon(1S),''
Phys.\ Rev.\ Lett.\  {\bf 92}, 222002 (2004)
[arXiv:hep-ex/0307034]; \\
%%CITATION = HEP-EX 0307034;%%
%\cite{Glenn:1998bd}
%\bibitem{Glenn:1998bd}
S.~Glenn {\it et al.}  [CLEO Collaboration],
%``Upsilon dipion transitions at energies near the Upsilon(4S),''
Phys.\ Rev.\ D {\bf 59}, 052003 (1999)
[arXiv:hep-ex/9808008].
%%CITATION = HEP-EX 9808008;%%

\bibitem{pythia}
%\cite{Sjostrand:2000wi}
%\bibitem{Sjostrand:2000wi}
  T.~Sjostrand, P.~Eden, C.~Friberg, L.~Lonnblad, G.~Miu, S.~Mrenna and E.~Norrbin,
  %``High-energy-physics event generation with PYTHIA 6.1,''
  Comput.\ Phys.\ Commun.\  {\bf 135}, 238 (2001)
  [arXiv:hep-ph/0010017].
  %%CITATION = HEP-PH 0010017;%%

\bibitem{comphep}
%\cite{Pukhov:1999gg}
%\bibitem{Pukhov:1999gg}
  A.~Pukhov {\it et al.},
  %``CompHEP: A package for evaluation of Feynman diagrams and integration  over
  %multi-particle phase space. User's manual for version 33,''
  arXiv:hep-ph/9908288.
  %%CITATION = HEP-PH 9908288;%%

\bibitem{tauola}
%\cite{Jadach:1993hs}
%\bibitem{Jadach:1993hs}
  S.~Jadach, Z.~Was, R.~Decker and J.~H.~Kuhn,
  %``The tau decay library TAUOLA: Version 2.4,''
  Comput.\ Phys.\ Commun.\  {\bf 76}, 361 (1993).
  %%CITATION = CPHCB,76,361;%%

\bibitem{babartdr}
%\cite{Boutigny:1995ib}
%\bibitem{Boutigny:1995ib}
D.~Boutigny {\it et al.}  [BABAR Collaboration],
%``BaBar technical design report,''
SLAC-R-0457
%\href{http://www.slac.stanford.edu/spires/find/hep/www?r=slac-r-0457}{SPIRES entry}

\bibitem{y3s1spipi}
%\cite{Chakravarty:1992zt}
%\bibitem{Chakravarty:1992zt}
  S.~Chakravarty and P.~Ko,
  %``On the pi pi Spectrum in Upsilon (3S) $\to$ Upsilon (1S) pi pi,''
  Phys.\ Rev.\ D {\bf 48}, 1205 (1993);\\
  %%CITATION = PHRVA,D48,1205;%%
%\cite{Chakravarty:1992tc}
%\bibitem{Chakravarty:1992tc}
  S.~Chakravarty, S.~M.~Kim and P.~Ko,
  %``Is Upsilon (3S) a pure S wave?,''
  Phys.\ Rev.\ D {\bf 48}, 1212 (1993)
  [arXiv:hep-ph/9301249]; \\
  %%CITATION = HEP-PH 9301249;%%
%\cite{Chakravarty:1993er}
%\bibitem{Chakravarty:1993er}
  S.~Chakravarty, S.~M.~Kim and P.~Ko,
  %``Final state pi pi interactions in Upsilon (3s) $\to$ Upsilon (1s) pi pi,''
  Phys.\ Rev.\ D {\bf 50}, 389 (1994)
  [arXiv:hep-ph/9310376];\\
  %%CITATION = HEP-PH 9310376;%%
%\cite{Mannel:1995jt}
%\bibitem{Mannel:1995jt}
  T.~Mannel and R.~Urech,
  %``Hadronic decays of excited heavy quarkonia,''
  Z.\ Phys.\ C {\bf 73}, 541 (1997)
  [arXiv:hep-ph/9510406].
  %%CITATION = HEP-PH 9510406;%%

%\cite{Guo:2004dt}
\bibitem{Guo:2004dt}
  F.~K.~Guo, P.~N.~Shen, H.~C.~Chiang and R.~G.~Ping,
  %``Heavy quarkonium pi+ pi- transitions and a possible b anti-b q anti-q
  %state,''
  arXiv:hep-ph/0410204.
  %%CITATION = HEP-PH 0410204;%%

%\cite{Budnev:1974de}
\bibitem{Budnev:1974de}
V.~M.~Budnev, I.~F.~Ginzburg, G.~V.~Meledin and V.~G.~Serbo,
%``The Two Photon Particle Production Mechanism. Physical Problems.
%Applications. Equivalent Photon Approximation,''
Phys.\ Rept.\  {\bf 15}, 181 (1974).
%%CITATION = PRPLC,15,181;%%

%\cite{Brambilla:2004wf}
\bibitem{Brambilla:2004wf}
N.~Brambilla {\it et al.},
%``Heavy quarkonium physics,''
arXiv:hep-ph/0412158.
%%CITATION = HEP-PH 0412158;%%


\end{thebibliography}
\end{document}